# Superconducting Quantum Computing Without Entanglement?

Alan M. Kadin, *Senior Member, IEEE* and Steven B. Kaplan, *Senior Member, IEEE*

*Abstract*—In recent years, quantum computing has promised a revolution in computing performance, based on massive parallelism enabled by many entangled qubits. Josephson junction integrated circuits have emerged as the key technology to implement such a universal digital quantum computer. Indeed, prior experiments have demonstrated simple Josephson qubit configurations with quantized energy levels and long coherence times, which are a necessary prerequisite for a practical quantum computer. However, these quantized states do not directly prove the presence of entanglement or macroscopic superposition, which are essential for the superior speed of such a digital quantum computer. On the contrary, an alternative realistic foundation for quantum mechanics has recently been proposed, with coherent transitions between quantized states, but *without* entanglement. It is suggested that the observations to date on superconducting circuits may be consistent with this quantum realism. A new experiment is proposed that may test whether superconducting quantum circuits can exhibit quantized states without macroscopic entanglement or superposition. Specifically, a flux qubit (a bi-stable SQUID) may be configured with a resonant input line for excitation and a single-flux-quantum (SFQ) output line for simultaneous direct measurement of quantized energy and flux states, which are incompatible measurements in standard quantum theory. Such an observation could undermine the assumptions of superposition and entanglement, bringing into question the foundation and the ultimate performance of a universal digital quantum computer. In contrast, this realistic quantum model may be compatible with localized quantum transitions in networks of superconducting circuits, which form the basis for an alternative technology of an analog quantum computer that implements quantum annealing. Such an analog computer has been fabricated and demonstrated successfully in a medium-scale superconducting circuit implementation. While this may not provide the revolutionary performance promised for an entanglement computer, it may still permit rapid simulation of model quantum systems, as well as certain classical optimization problems. Other experimental implications of this realistic quantum picture are also discussed.

## I. Introduction

Quantum computing applies the formalism of quantum mechanics, developed for describing the behavior of electrons in atoms, to electronic devices on the macroscopic scale [1]. Electrons are certainly waves, and standing waves on the atomic scale (i.e., wave superpositions) are essential to understanding atomic bonding and energy gaps. It is not quite so obvious that composite objects like Josephson junctions should be described in the same way. However, in orthodox quantum theory, any object on any scale can be properly represented by an abstract wavefunction in Hilbert space, provided only that it does not exchange energy with its environment on a relevant timescale. While this was widely debated in the early development of quantum mechanics, it is now taken as a matter of faith, and never questioned. By the same token, early critics of orthodox quantum theory pointed out the absence of a consistent realistic picture of quantum waves, leading to physical and logical paradoxes that have never really been resolved.

The present paper is dedicated to the unorthodox proposition that these paradoxes represent a genuine problem [2], which is not merely a question of "interpretation". Any physically distinct representation will have real experimental implications, which can be tested using carefully designed experiments and modern instrumentation. Superconducting devices may be ideal for such tests, since they provide both quantum-limited systems and electronically accessible measurements.

Specifically, a superconducting loop can carry a circulating current associated with a quantized magnetic flux (Fig. 1). It can also be configured to carry a circulating current of the opposite sign. This is analogous to an electron in an atom, which in a p-orbital can carry angular momentum in either sense around a given axis. In the atom, one may also construct a rotational standing wave, composed of a sum of waves rotating in both directions. This is a $p_z$-state with zero angular momentum and directional lobes. Can one analogously have a superposition state of opposite currents in a superconducting loop, comprising a "flux qubit" [3]? If the states going left and right are $\Psi_L$ and $\Psi_R$, the states $\Psi_\pm$ are

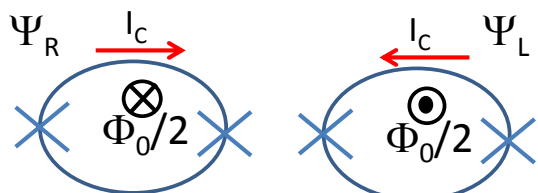

Fig. 1. A SQUID may be designed to be bi-stable, i.e., to have two stable current configurations, corresponding to current and field going in opposite directions. According to the orthodox quantum theory, this "flux qubit" may be in a quantum state that is a superposition of the two realistic basis states. On the contrary, it is argued here that no such superposition states exist, and that a new experiment can provide direct evidence on this question.

Manuscript received August 11, 2014.
A. Kadin is a consultant based in Princeton Junction, NJ 08550 USA, e-mail: amkadin@alumni.princeton.edu.
S. Kaplan is a consultant based in Yorktown Heights, NY 10598 USA, email: stevebkaplan@optonline.net.

defined by

$$\Psi_\pm = (\Psi_L \pm \Psi_R)/\sqrt{2} \qquad (1)$$

From a realistic point of view, this is nonsense. At any one time, the electrical current in a macroscopic loop flows in one direction or the other, not in both at the same time. Yet that is precisely what is claimed in the theory of quantum computing. Furthermore, it is widely believed that such superposition states have been observed experimentally [4,5]. It is argued here that only energy quantization has really been observed, which may not require such superposition states. Further, a new experiment is proposed that may distinguish a superposition state from one with definite flux.

A closely related concept is quantum entanglement [6,7], in which a quantum state $\Psi_{tot}$ of two interacting electrons $\Psi_1(A)$ and $\Psi_2(B)$ is a superposition of the product of the two states with their states A and B exchanged, taking the form

$$\Psi_{tot} = [\Psi_1(A)\,\Psi_2(B) - \Psi_1(B)\,\Psi_2(A)]/\sqrt{2}. \qquad (2)$$

This construction initially provided the explanation for the Pauli exclusion principle in atoms, but subsequently became the basis for a general Hilbert-space formalism of interacting quantum states. Such entangled states also provide the basis for the theory of quantum computing. The concept of quantum entanglement was criticized early in the history of quantum mechanics by Erwin Schrödinger, as epitomized in his famous "cat paper." Schrödinger presented a coupled quantum state of a radioactive atom and a cat, and argued that one could construct a state in which the cat was in a linear combination of being dead and alive. This, he argued, showed that the orthodox theory of quantum mechanics was logically inconsistent. Nevertheless, it is now universally believed that such "Schrödinger cat states" are proven facts. On the contrary, it is argued here that entangled states are incompatible with any realistic quantum theory, and do not exist in nature.

Recently, an alternative quantum picture was proposed [2] in which primary quantum fields (electrons, photons, quarks) form real-space wave packets with quantized spin, without point particles or fundamental uncertainty. Composites of these quantized wave packets may follow particle-like trajectories but are not themselves waves, and are not subject to superposition. Within this picture, composites exhibit quantization of energy and angular momentum by inheriting these properties from their constituents. So quantum mechanics is really a mechanism for a world of microscopic continuous fields to condense into a world of discrete particles, rather than a universal theory of all matter. A further implication of this picture is that a quantum transition represents a continuous reconfiguration of the primary fields in such a composite, mediated by a photon with spin $S=\hbar$. The Hilbert space formalism of the orthodox theory does not properly describe such composites and their transitions.

## II. Quantum Computing and Entanglement

Quantum computing is being developed because it promises to enable solutions to computing problems that cannot efficiently be solved by any classical computer or even a large parallel array of such computers [1]. One such difficult problem is the factoring of very large integers, the difficulty of which forms the basis of modern encryption technology. It would be of great interest to many people (especially those in spy agencies) if an efficient way to decode encrypted messages were possible. Digital quantum computing promises to do exactly this, using Shor's algorithm.

In essence, quantum computing enables the equivalent of exponentially large parallelism with a much smaller array of devices. Such a claim is quite remarkable, and should not be accepted without clear experimental verification. In its simplest form, the argument for such exponential parallelism builds on the Hilbert space formalism for which quantum superposition and entanglement are essential elements. Consider first a classical N-bit computer processor, with hardware comprising N bit slices operating in parallel. This can represent $2^N$ integers, but a given computation can only follow one of these integers at a given time. Now consider a quantum computer with N qubits. According to the orthodox theory, each qubit represents a superposition of two states (a two-dimensional Hilbert space), and a quantum processor can follow the evolution of both of these in parallel. Further, two coupled qubits will be entangled, which multiplies the dimension of the Hilbert space, creating 4 basis states that can be followed in parallel. Generalizing to N entangled states, a quantum processor can follow $2^N$ states in parallel. Taking an example of N=300, this represents a parallel speedup of $2^{300} \sim 10^{90}$. So a 300-qubit quantum processor could have the equivalent computing power of $10^{90}$ classical bit slices operating in parallel. The latter is clearly impossible.

In contrast, within the realistic quantum picture of [2], there is no quantum entanglement, no quantum superpositions of composite systems, and the Hilbert space expansion does not describe real N-qubit systems. In that case, there would be no exponential parallelism, and no corresponding speedup to solve such problems as factoring large numbers. Would there be anything left for quantum computing, using super-conducting devices or other elements?

In this regard, it is useful to examine a very different approach to quantum computing which has received much less attention in the academic research community. This is adiabatic quantum computing, which really provides a form of analog computing rather than universal digital computing [8]. A superconducting electronic processor of this class [9,10] has been developed by D-Wave Systems in Canada, built around a two-dimensional array of 512 Nb flux qubits cooled down to ~ 10 mK. This was designed as a quantum annealing processor, which maps onto the 2D Ising model of magnetic interactions. This is an analog computer that models a magnetic phase transition, but also can be used to solve other classical and quantum optimization problems that map onto the Ising model. This can be compared to classical simulated annealing, where a model crystal is gradually cooled toward T=0, to obtain the ground state energy and configuration. But a crystal can "freeze in" local defects, which can have extremely slow relaxation times as T→0. Quantum annealing can relax such defects via local quantum tunneling, enabling more rapid convergence to the ground state. This requires local quantum transitions, but not necessarily quantum superposition and entanglement. This system is currently being tested in multiple laboratories, and seems to exhibit some quantum speedup compared to classical thermal annealing [11]. While D-Wave is moving toward increasing the number N of qubits, it would appear that the system performance scales much more slowly than the exponential enhancement predicted for digital quantum computing. Still, this superconducting *analog* quantum computing approach may have a promising future for solving certain optimization and quantum models, even if the predicted revolutionary performance for digital quantum computing should turn out to be invalid.

### III. PROPOSED FLUX QUBIT EXPERIMENT

In order to address the question of whether a superconducting qubit can exist in a superposition state, an experiment is proposed based on a flux qubit structure, as shown schematically in Fig. 2. The flux qubit is based on a two-junction SQUID, with loop inductance $L \sim 1.5\Phi_0/I_0$ for bi-stable operation (to store a single flux quantum $\Phi_0 = h/2e$), where $I_0$ is the critical current of the junctions. This has a signal coupled inductively to the input, and its state may be measured by coupling to a second SQUID detector/amplifier on the right, as shown in Fig. 2a. The intent is to demonstrate quantum coherence in the qubit, while at the same time monitoring the flux state of the qubit.

According to the orthodox quantum theory, the former requires that the qubit be in an energy eigenstate, while the latter requires that the qubit be in a flux eigenstate. The energy eigenstates are generally believed to be superposition states $\Psi_\pm = (\Psi_L \pm \Psi_R)/\sqrt{2}$, in contrast to the flux eigenstates $\Psi_L$ and $\Psi_R$. In orthodox theory, these are incompatible measurements. In contrast, in the realistic theory, an energy transition would be associated with a flux transition, mediated by a photon that transfers angular momentum $\hbar$ (and flux h/2e).

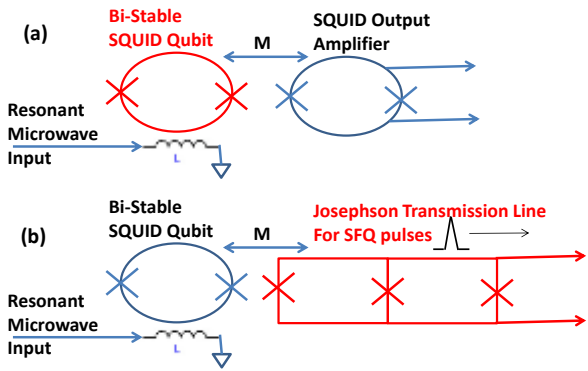

Fig. 2. Concept of proposed experiment involving a flux qubit structure with a resonant input to probe quantized energy and a flux output to probe the flux state. (a) A SQUID output amplifier designed to measure directly and continuously the flux state of the qubit. (b) An improved output designed to measure a change in the flux state as a single-flux-quantum (SFQ) pulse, with less back-action on the qubit.

But there are important problems associated with the simple output circuit of Fig. 2a, focusing on the back-action of the measurement SQUID (right) on the flux qubit (left). A classic SQUID used as a detector or amplifier is operated at $I > 2I_0$ in the V>0 state. This requires a non-hysteretic I-V curve, which in turn requires using junctions that are critically damped, typically using an external resistive shunt, so that Q ~ 1. In contrast, a SQUID for a flux qubit must be a high-Q device in order to attain the quantum coherence associated with quantum transitions, so that the qubit junctions must be unshunted and cooled to a very low temperature.

It is also important to appreciate that a damped SQUID in the voltage state generates a rapid sequence of single-flux-quantum (SFQ) pulses, which create broadband noise that can couple back to the qubit, destroying the quantum coherence. For this reason, several experiments [12,13] have been carried out using an underdamped (unshunted) SQUID, which is itself a resonant system rather than a linear amplifier. If the critical current of the output SQUID is exceeded, a hysteretic output signal is triggered, ending the experiment. Other approaches have used measurements that may provide continuous monitoring of the system [14,15], but in a way that only indirectly provides information on the flux state of the qubit. Although the analyses concluded that quantum superposition was indeed present, a more direct measurement would be highly desirable.

A possible way to obtain a direct measurement without destroying quantum coherence is indicated in Fig. 2b. Here, rather than an output SQUID operating in the voltage state, the output flux is coupled to the input SQUID of a Josephson transmission line (JTL), with propagates a single-flux-quantum (SFQ) pulse down the line. The junctions in the JTL are always in the zero-voltage state, except during the ~1 ps when an SFQ pulse is generated. This should generate very little back-action on the qubit. Furthermore, the coupling parameters may be designed so that the damping effect of the nearest JTL junction may be minimized. The general concept of using SFQ circuits to read out a flux qubit was described some years ago by Feldman and Bocko [16,17], but evidently this approach was not fully followed through. It may be time to re-evaluate such an SFQ approach, using modern high-quality Josephson junctions.

### IV. SUPERPOSITION IN ELECTRONS AND PHOTONS

The analysis above has addressed the question of superposition of macroscopic quantum states, but there are also serious questions about microscopic superposition as well. These too may be subject to direct testing in experiments that should be straightforward using modern instrumentation. Although this is somewhat outside the subject of the present paper, the examples of electron spin and photon polarization represent paradigms for quantum measurement as well as for qubits in quantum computing. The proposed experiments have important implications for the foundations of quantum mechanics.

Specifically, orthodox theory states that an electron in a magnetic field may be either spin up, spin down, or any linear combination of these two. On the contrary, the spin-quantized picture asserts that the electron must be either spin up or spin down, and never in a superposition. Similarly, orthodox theory states that a single photon may be in a left circularly polarized or right circularly polarized state (left or right CP), or any linear combination of these two. The linear combinations enable a single photon to be in a state of linear or elliptical polarization (LP or EP). On the contrary, the spin-quantized picture asserts that a single photon must always be CP with a spin $S = \pm \hbar$, and can never be LP or EP. Within this picture, one can only construct LP or EP fields by a sum of multiple CP single photons.

Consider first the electron spin problem, and revisit the classic Stern-Gerlach (SG) experiment (1922), in which an atomic beam of univalent atoms is sent into an inhomogeneous magnetic field. The incoming beam is split into two output sub-beams with spin up or spin down. This is predicted by both of these two quantum pictures. However, a two-stage SG experiment (Fig. 3), in which one sub-beam is further sent into a second SG magnet rotated by an angle θ, could distinguish these two approaches. This is a standard textbook example, which was introduced as a thought-experiment in *The Feynman Lectures on Physics* [18]. The predicted result is that the beam intensities of the second selector should go statistically as $\cos^2\theta$ and $\sin^2\theta$. In contrast, in the spin-quantized picture [2], the result should be 1 and 0, with no statistical variations. The spins in the excited state will adiabatically rotate to follow the direction of the rotating field that they see. This experiment has apparently never been done (as Feynman admitted), but could be easily tested using modern atomic beam equipment.

Now consider the case of linearly polarized single photons, which are universally believed to have been measured in many quantum optics experiments, including those used to demonstrate quantum entanglement. However, most conventional single-photon detectors (such as photomultipliers or avalanche photodiodes) are really event detectors without energy resolution, and cannot distinguish a single photon from two simultaneous photons. But state-of-the-art superconducting single-photon detectors based on transition-edge sensors (TES) can measure total deposited energy with less than 1-eV resolution, and have better than 90% quantum efficiency [19,20,21]. This is relevant because in the spin-quantized picture [2], a "single photon" should really be the sum of a left CP and a right CP photon. Recent experiments with attenuated pulsed laser beams have shown that such a TES detector can identify the number of photons in each pulse, as indicated in the lower figure in the block diagram of Fig. 4, which shows a statistical distribution with 1, 2, 3, and 4 photons. It is suggested [22] that if the same experiment were done with a linear polarizer in the beam path, the odd photon counts would be expected to drop out (according to the spin-quantized picture), leaving only even counts (2,4). In contrast, the orthodox picture permits LP single photons, so both odd and even counts would be maintained. Such a simple, direct experiment has not yet been reported.

## V. CONCLUSION

Superconducting quantum computing promises to provide a revolution in computer performance, by enabling exponential parallelism that cannot be matched by any conceivable array of classical computers. This argument is based on fundamental concepts of entanglement and superposition that have been questioned by a novel realistic quantum picture. An experiment is proposed, using a superconducting flux qubit structure and a single-flux-quantum output, which may directly test the principle of superposition in these devices. Without such superposition, the exponential speedup may not be present, undermining the major motivation for quantum computing. However, even if this were the case, the alternative quantum annealing approach to an analog quantum computer should still be able to provide quantum improvement to calculating certain optimization problems. This realistic quantum picture also suggests experiments that would directly test superposition in electron spin and photon polarization, including an experiment using sensitive superconducting single-photon detectors.

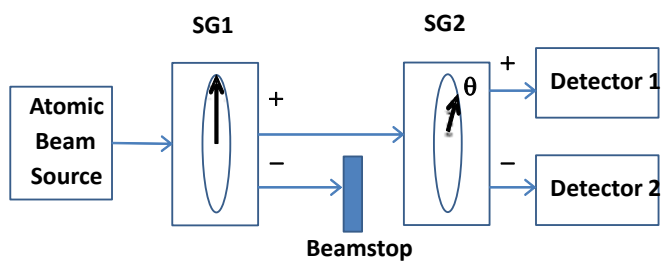

Fig. 3. Block diagram of a two-stage Stern-Gerlach (SG) experiment, where the angle of the magnetic field in the 2$^{nd}$ stage is rotated by an angle θ. This experiment is a standard textbook example of quantum measurement, but apparently has never actually been done. The prediction of orthodox theory is that the beam intensities at the output of the 2$^{nd}$ stage should go as $\cos^2\theta$ and $\sin^2\theta$, while in the heterodox spin-quantized picture [2], the intensities should go as 1 and 0, with no statistics.

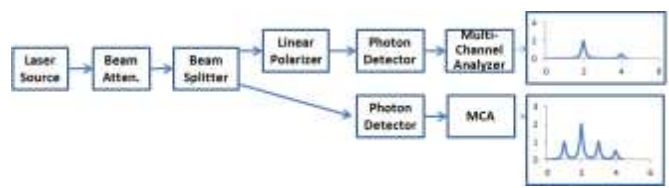

Fig. 4. Block diagram of proposed experiment [22] to test whether linearly polarized "single photons" are really photon pairs, using a superconducting single photon detector with fine energy resolution and high quantum efficiency. Without a linear polarizer [19,20,21], the number of photons in an attenuated pulse may be N = 1, 2, 3, or 4. With the polarizer in place, the odd counts (including N = 1) should drop out, according to the spin-quantized quantum picture [2].